\documentclass[preprint,12pt,authoryear]{elsarticle}

\usepackage{algorithm}
\usepackage{algpseudocode}

\usepackage{graphicx,epsfig} 
\usepackage[utf8]{inputenc}
\usepackage[T1]{fontenc}
\usepackage{amsmath,amssymb,amsfonts, bbm}
\usepackage{amsthm}
\usepackage{xargs}                      
\newcommand{\inR}{\in\mathbb{R}^}
\newcommand{\R}{\mathbb{R}}
\newcommand{\nn}{f_{NN}}
\newcommand{\ctg}{cost-to-go }
\newcommand{\psd}{\succcurlyeq} 
\newcommand*\figref{Fig.~\ref} 
\usepackage[pdftex,dvipsnames]{xcolor}  
\usepackage{babel,csquotes,xpatch}
\newtheorem{theorem}{Theorem}
\newproof{pf}{Proof}

\usepackage{caption}
\usepackage{subcaption}
\usepackage{comment}

\usepackage{hyperref}
 \usepackage{xurl}
\hypersetup{
    colorlinks=true,
    linkcolor=blue,
    filecolor=magenta,      
    urlcolor=cyan,
    }
\allowdisplaybreaks 

\usepackage{microtype}

\begin{document}
\begin{frontmatter}

\author[inst1]{Evren Mert Turan}

\affiliation[inst1]{organization={Department of Chemical Engineering, Norwegian University of Science and Technology (NTNU)},
            addressline={Sem Sælandsvei 4}, 
            city={Trondheim},
            postcode={7491}, 
            country={Norway}}
\author[inst1]{Zawadi Mdoe}

\author[inst1]{Johannes Jäschke}\ead{johannes.jaschke@ntnu.no}
\title{Learning a convex cost-to-go for single step model predictive control}

\begin{abstract}
{\color{black}
For large uncertain systems, solving model predictive control problems online can be computationally taxing. Using a shorter prediction horizon can help, but may lead to poor performance and instability without appropriate modifications. This work focuses on learning convex objective terms to enable a single-step control horizon, reducing online computational costs.
We consider two surrogates for approximating the cost-to-go: (1) a convex interpolating function and (2) an input-convex neural network. Regardless of the surrogate choice, its behavior near the origin and its ability to describe the feasible region are crucial for the closed-loop performance of the new MPC problem. We address this by tailoring the surrogate to ensure good performance in both aspects.
We conclude with numerical examples, in which we compare the convex surrogates to using a standard neural network in the objective, solely using an LQR cost-to-go, and to using a neural network to learn a control policy. The proposed approaches are shown to achieve better performance with less data.
}

\end{abstract}


\end{frontmatter}


\section{Introduction}
\label{cnv-cost-sec:intro}
Model predictive control (MPC) is an optimisation-based control method, in which a control action that minimises some objective while satisfying constraints is found by use of a model that predicts the (short-term) response of a system given the current state. An optimisation problem has to be solved to find the control action, which can be computationally infeasible for large problems or fast systems, especially when considering a robust MPC formulation. The computational burden can be reduced by considering a shorter horizon (by reducing the problem size); however, this can often excessively deteriorate the controller performance. In this work, we aim to reduce the computational burden of (robust) MPC by learning a convex control objective that allows the use of a prediction and control horizon of one.

Various approaches have been considered to reduce the online computational delay of MPC. One approach is to compute the explicit feedback control law that is implicitly defined by the MPC problem. This can be done for a standard linear MPC problem by solving a multi-parametric programming problem \citep{Bemporad2002}. This method is limited to relatively small-scale problems as the online computational requirements grows exponentially with the problem size. One can instead consider finding a compact parameterisation of the control policy by a neural network, trained in either an imitation learning \citep{Karg2020,kumar2021industrial} or optimize-and-learn framework \citep{Turan2023-nn-policy-jpc}. {\color{black}However, these policies can be difficult to adjust online, when for example a constraint changes.}

An alternative approach is to consider solving a smaller problem online. However, a smaller problem does not necessarily yield an effective control policy -- a naive implementation of an MPC problem with a horizon of 1 will often give bad results. 
However, if one uses a problem with a short horizon, with an appropriately designed cost-to-go term, there is no loss in performance when using the shorter horizon.

{\color{black}Two approaches that can be used to find such a cost-to-go term are} inverse optimal control and approximate dynamic programming. In inverse optimal control
a data-set of state-input pairs from a controller or an expert is fitted to a simple MPC controller by learning a value function, such that the simple MPC controller is approximately optimal  \citep{keshavarz2011}.
In approximate dynamic programming, the value function is approximated (commonly iteratively) based on some metric, e.g. \citet{wang2009performance} solved a semi-definite problem offline to find a convex quadratic value function to approximate the cost-to-go.

{\color{black}In the present paper, we consider learning (offline) a convex surrogate of the \ctg of a convex (robust) MPC problem with the primary aim of reducing the online computational cost of the MPC problem.  We emphasise that this problem class includes linear MPC (the most commonly implemented MPC) and more general problems, e.g., a problem with convex (but potentially non-linear) state, cost and input inequality constraints, and linear dynamics is convex. One can also construct convex MPC problems through Disciplined Convex Programming \citep{grant2006disciplined} to ensure convexity. 

If the original problem is convex, then use of a convex surrogate has several clear benefits. Firstly using a convex surrogate maintains convexity. This avoids the standard difficulties of optimising over general surrogates (see \citet{ceccon2022omlt} and the references therein).} Furthermore, the restriction of the surrogate to be convex can be regarded as a form of regularisation thus avoiding the difficulty of tuning the training problem to avoid over-fitting. {\color{black} Our numerical results imply that this restriction improves the data efficiency of learning the surrogate. Lastly, we would like to note that despite convexity, convex MPC problems can still be challenging to solve in real-time due to the model size, e.g. see \citet{kumar2021industrial}, or when considering uncertainty.}

{\color{black}We} focus on two convex surrogates: (1)
an interpolating convex function given as the solution of a convex function and (2) an input-convex neural network (ICNN), which is a network that is non-convex to train but convex to optimise over. In contrast to learning a control policy, this approach is more flexible as changes to problem data can be incorporated by partially lengthening the control horizon and updating the optimisation problem with the new data.

Related approaches were reported in \citep{Abdufattokhov2021,seel2022convex,orrico2024building}, but our approach differs as follows: In \citet{Abdufattokhov2021}, a quadratic cost function is parametrised by a neural network, i.e. the quadratic term is $x^TL^TLx$, and the neural network learns $L$,  such that the approximation error of the parametric quadratic objective and the \ctg is minimised. In contrast, we consider the direct approximation of the {\color{black}cost-to-go. As our surrogates are convex, they can be effectively optimised over.} In \citet{seel2022convex}, the parameters of an ICNN in the objective are adjusted online in a reinforcement learning scheme to give good controller performance. In contrast, we are interested in training a convex approximation \textit{offline} to give good performance when predicting only a single step ahead. Although one of the approximations we consider is an ICNN the problem formulation and training differ. Lastly, and most similar to the approach proposed in this work, in \citet{orrico2024building} a neural network is used to learn the \ctg term of an MPC problem.  {\color{black} However, in this work we propose formulations for learning a convex surrogate instead of a generic non-linear surrogate. When specifically considering neural networks, we demonstrate that even in a low-dimensional example, if the original problem is convex, using an ICNN  significantly reduces the amount of training data needed to achieve reasonable error in the control output. This improvement in data is important as generating sufficient amounts of training data can be challenging for larger problems, e.g. see the discussion in \citet{kumar2021industrial}. Additionally, naive optimisation over general non-linear surrogates can lead to computationally challenges \citep{ceccon2022omlt}.} 

The paper is organised as follows: Section \ref{cnv-cost-sec: problem-form} briefly states the problem formulation, Section \ref{cnv-cost-sec: learning the ctg} details how the \ctg is approximated, Section \ref{cnv-cost-sec: approximating=cnvx-fns} introduces the two choices of convex surrogates (interpolating convex functions and input-convex neural networks), Section \ref{cnv-cost-sec: results} numerically demonstrate the proposed approach and Section \ref{cnv-cost-sec: conc} concludes the paper. 
\section{Problem formulation}\label{cnv-cost-sec: problem-form}

Consider a convex MPC problem {\color{black}with linear dynamics}\footnote{{\color{black} Note that the presented approach could also be applied to any convex MPC problem, but convexity can be difficult to ensure with equality constraints. The approach could also be applied to general non-linear MPC problems, however some of the beneficial properties of the approach would be lost.}}:
\begin{subequations}\label{cnv-cost-eqnstandard-mpc}
\begin{align}
    \mathcal{V}_N(\hat{x}) &= \min_{u,x}\:\: \sum_{k=0}^{N-1} l_k(x_k,u_k)+V_t(x_N)\\
    x_{k+1} &= Ax_k+Bu_k,\qquad k=0,\dots,N-1 \label{cnv-cost-eqnstandard-mpc nom-dynamics}\\
    u_k &\in \mathcal{U}_k,\qquad k=0,\dots,N-1\\
    x_0 &= \hat{x},\: x_k \in \mathcal{X}_k,\qquad k={\color{black}0},\dots,N 
\end{align}
\end{subequations}
where the system is to be regulated to the origin, and where $N$ is the prediction horizon, $x_k\in\mathcal{X}_k\subseteq\R^{n_x}$ are the states, $u_k\in\mathcal{U}_k \subseteq \R^{n_u}$ are the control inputs, $\hat{x}\inR {n_x}$ is the initial condition, and estimate of the current state of the process, $k$ indexes the discrete time model, $l_k$ is a convex stage cost, $V_t$ is a convex terminal cost,  $\mathcal{V}_N$ is the optimal value function with horizon $N$, and $\mathcal{U}_k\subset\R^{n_u}$ and $\mathcal{X}_k\subset\R^{n_x}$ {\color{black}are convex constraint sets}.

{\color{black}As \eqref{cnv-cost-eqnstandard-mpc} is the minimization of a convex objective over a convex set, and $\hat{x}$ enters \eqref{cnv-cost-eqnstandard-mpc} linearly,  the value function $\mathcal{V}_N$ is  a convex function of $\hat{x}$.} Typically $l_k$ and $V_t$ are strictly convex quadratic functions:
\begin{subequations}\label{cnv-cost-eqnquadratic-costs}
\begin{align}
l(x_k,u_k)&=x^T_kQx_k+u^T_kRu_k\\
V_t(x_N) &= x_NQ_fx_N
\end{align}
\end{subequations}
where $Q,\ R$ and $Q_f$ are symmetric positive definite matrices.  {\color{black} If $l_k$ and $V_t$ are defined as above, then $V_t$ is often chosen as the value function of the unconstrained infinite horizon control problem corresponding to  \eqref{cnv-cost-eqnstandard-mpc}, i.e. the linear quadratic regulator (LQR).}

\subsection{Multistage MPC}
Problem \eqref{cnv-cost-eqnstandard-mpc} is a {nominal} MPC problem as it assumes that the system is perfectly described by \eqref{cnv-cost-eqnstandard-mpc nom-dynamics}. Realistically, this is not the case due to uncertainty. 
This uncertainty can either be ignored (in which case we solve the nominal problem, \eqref{cnv-cost-eqnstandard-mpc}) or can be incorporated in some robust or probabilistic framework. We present our results for the multistage MPC framework \citep{Lucia2013,Scokaert1998}, although the main results can be applied to other {\color{black}convex} formulations that take into account uncertainty.

In a multistage MPC problem, we consider a similar system to \eqref{cnv-cost-eqnstandard-mpc} but explicitly consider uncertainty by including predictions corresponding to $S$ different scenarios in which the process dynamics are different:
\begin{subequations}\label{cnv-cost-eqnms-mpc}
\begin{align}
    \mathcal{V}^{MS}_N(\hat{x}) &= \min_{u,x}\:\: \sum_{s=1}^{S}{w_s}\mathcal{V}_{N,s}(\hat{x})\\
   \mathcal{V}_{N,s}(\hat{x})&= \left(\sum_{k=0}^{N-1} l_k(x_{k,s},u_{k,s})+V_t(x_{N,s})\right) s=1,\dots,S\\
    x_{k+1,s} &= A_{k,s}x_{k,s}+B_{k,s}u_{k,s} + d_{k,s},\nonumber\\&\ k=0,\dots,N-1,\  s=1,\dots,S\\
    u_{k,s} &\in \mathcal{U}_k,\qquad k=0,\dots,N-1\  s=1,\dots,S\\
     x_{0,s} &= \hat{x},\ x_{k,s} \in \mathcal{X}_k,\qquad  k={\color{black}0},\dots,N \   s=1,\dots,S\label{eq:  constraint sets}\\
    x_{k,s_1} &= x_{k,s_2} \Rightarrow   u_{k,s_1} = u_{k,s_2},\   s_1, s_2 =1,\dots,S \label{cnv-cost-eqnnon-antici} 
\end{align}
\end{subequations}
where $s$ indexes the different scenarios, and the objectives of the different scenarios are weighted by $w_s>0$ with $\sum_sw_s=1$. 
$A_{k,s}$, $B_{k,s}$, and $d_{k,s}$ are realisations of the stochastic or uncertain parameters which are typically decided upon offline. Regardless of the parameter values in the different scenarios, 
$\mathcal{V}^{MS}_N$ remains a convex function. Constraint \eqref{cnv-cost-eqnnon-antici} is a non-anticipativity constraint that enforces the control action of two scenarios to be equal at a time point if the states of two scenarios are equal at the same time point. 

\begin{figure}
     \centering
     \includegraphics[width=0.99\linewidth]{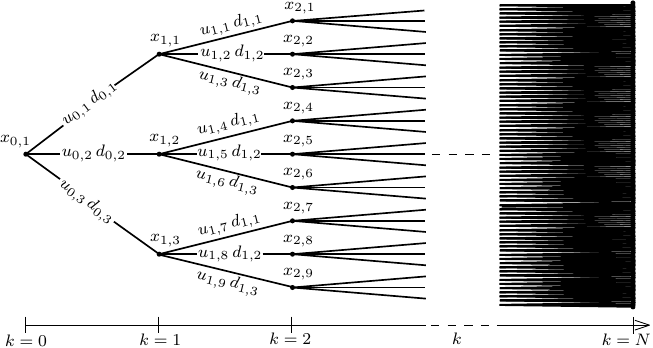}
     \caption{Fully branched scenario tree.}
     \label{fig:scenariotreefullybranched}
\end{figure}

The uncertainty realisations can be represented by a scenario tree, with branching representing a potential change in the uncertainty realisation.  A tree is called fully branched if branching occurs until the end of the prediction horizon $N$, i.e. all possible system evolutions for a finite number of parameter realisations are considered. 
A scenario tree for a system with three potential values for $d$ is shown in \figref{fig:scenariotreefullybranched}. 
The size of the fully-branched scenario tree, and consequently the multistage MPC problem, grows exponentially with the horizon length and number of parameter values. Thus, even though the problem remains convex, it is often computationally infeasible to solve a large, fully-branched multistage MPC problem with a long horizon. {\color{black}To reduce complexity, one can heuristically consider only branching the tree for a shorter horizon, called the robust horizon.}

\subsection{Only a step-ahead}
Using Bellman's principle of optimality problems \eqref{cnv-cost-eqnstandard-mpc} and \eqref{cnv-cost-eqnms-mpc} may be reformulated as the single step problem with a horizon of 1. As  \eqref{cnv-cost-eqnstandard-mpc} is a special case of \eqref{cnv-cost-eqnms-mpc} we only show the reformulation of \eqref{cnv-cost-eqnms-mpc}:
\begin{subequations}\label{cnv-cost-eqn1stage-mpc}
\begin{align}
    \mathcal{V}^{MS}_N(\hat{x}) =  \min_{u_0,x_{1,s}}\:\:  
    &l_0(\hat{x},u_0)+\sum_{s=1}^{S}{w_s}(\mathcal{V}_{N-1}^{MS}(x_{1,s}))&\\
    x_{1,s} &= A_{0,s}\hat{x}+B_{0,s}u_{0} + d_{0,s}\qquad  s=1,\dots,S\\
    u_0 &\in \mathcal{U}_0,\  x_{1,s} \in \mathcal{X}_1\:\: s=1,\dots,S \label{eq: state-constraint-short}
\end{align}
\end{subequations}
Only a single step prediction into the future is used, however the section of the trajectory that is left out is captured by the embedded value function, $\mathcal{V}^{MS}_{N-1}$, which is called the cost-to-go. 
{\color{black}As the uncertainties are independent and not a function of the state, the same cost-to-go function, $\mathcal{V}^{MS}_{N-1}$, is evaluated for each scenario at $x_{1,s}$.}
The computational cost of solving a nominal MPC problem online could be greatly reduced if $\mathcal{V}^{MS}_{N-1}$ was explicitly known and could be evaluated easily, as the optimization problem {\color{black}is much smaller}.
This benefit is compounded when considering a fully branched multistage problem, as using a control horizon of 1 means that the problem size grows \textit{linearly} and not exponentially with the number of scenarios per branch point.

\section{Problem formulation}\label{cnv-cost-sec: learning the ctg}
In this section, we describe the design concerns of the convex surrogate for use in the MPC problem and the problem of learning feasibility. We would first make the following note with regard to the approximation error of the surrogate.  When finding a convex approximation, $\hat{\mathcal{V}}$ we are interested in using $\hat{\mathcal{V}}$ in  \eqref{cnv-cost-eqn1stage-mpc} to yield control actions that {\color{black} are equivalent to solving the full-horizon problem.

Therefore we can tolerate errors in the cost-to-go approximation if this does not change the optimal control actions, e.g. a constant bias will not affect the solution.}

\subsection{Design of a convex \ctg approximation}\label{subcnv-cost-sec: design-approximator}
\begin{figure}
    \centering
    \includegraphics[width=0.6\linewidth]{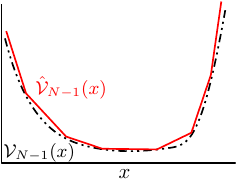}
    \caption{Illustration that a ``good" convex approximator (red line) of $\mathcal{V}_N$ (black line) can be a poor choice of function to minimize due to non-unique minimizers.}
    \label{fig:non-unique-min}
\end{figure}
To simplify the notation in this section  we use $\mathcal{V}(x)$ to denote $\mathcal{V}_{N-1}(x)$. Often ${\mathcal{V}}(x)$ is strictly convex , e.g. if $l_k$ and $V_t$ chosen as in \eqref{cnv-cost-eqnquadratic-costs}. However, the surrogate $\hat{\mathcal{V}}(x)$ may not be strictly convex resulting in the minimiser of \eqref{cnv-cost-eqn1stage-mpc} not being unique if the stage cost, $l$, is not strictly convex. 
This (unwanted) behaviour is typified by \figref{fig:non-unique-min} where a convex approximator approximates the region around the minimum by a line.  If \eqref{cnv-cost-eqn1stage-mpc}  has a non-unique minimum, then this implies that the controller using this approximation will show poor performance close to the origin and that the system may not be closed-loop stable.

To avoid this unwanted behaviour we can select $\hat{\mathcal{V}}$ as the sum of a convex term and a strictly convex term, e.g.
\begin{equation}
    \hat{\mathcal{V}}(x) = \hat{\mathcal{V}}_{sur}(x) + x^TPx,\:\: 0\le\hat{\mathcal{V}}_{sur}(x)\:\:  \label{cnv-cost-eqns-icnn}
\end{equation}
where $\hat{\mathcal{V}}$ is the approximate cost to go, $\hat{\mathcal{V}}_{sur}(x)$ is a convex surrogate that is already fitted, and $P\succ0$.
Although it is possible to optimise $\hat{\mathcal{V}}_{sur}$ and the entries of $P$ simultaneously, it is more practical to chose $P$ and then optimise for $\hat{\mathcal{V}}_{sur}$. 
Note that the resulting $\hat{\mathcal{V}}(x)$ is a strictly convex function with a unique minimizer.

When selecting $P$ one should ensure that $x^TPx$ is a lower bound of $\mathcal{V}^{MS}_N$, as otherwise $\hat{\mathcal{V}}_{sur}$ would need be non-convex for $\hat{\mathcal{V}}$ to be close approximation of $\mathcal{V}^{MS}_N$. If the stage cost is defined as in \eqref{cnv-cost-eqnquadratic-costs} then one can simply select $P=Q$. However, a more powerful idea is to select $P$ as the \ctg of the associated multistage linear quadratic regulator (LQR) problem., i.e. the multistage problem \eqref{cnv-cost-eqnms-mpc} with $\mathcal{X}_k=R^{n_x}$ and  $\mathcal{U}_k=R^{n_u}$. Importantly, one needs only to consider the multistage LQR problem \textit{without} additive disturbances, i.e. only parameteric uncertainty. 

{\color{black}
[Assumption 1] The $N$-stage multistage MPC problem \eqref{cnv-cost-eqnms-mpc}, denoted $\mathbb{P}^{MS}_N$, has the following properties:
\begin{itemize}
    \item problem $\mathbb{P}^{MS}_N$ is feasible and the optimal value function $\mathcal{V}^{MS}_N(x)$ is  finite for all $x\in\mathcal{X}_0$
    \item The stage and terminal costs \eqref{cnv-cost-eqnquadratic-costs} are strictly convex quadratic functions
    \item $w_{s,k}$ and $d_{s,k}$ be chosen such that $\sum_s^S w_sd_{s,k}=0$
\end{itemize}
These assumptions render problem $\mathbb{P}^{MS}_N$ a convex quadratic optimization problem with a unique minimizer. {\color{black} The last assumption is that the $d_{s,k}$'s weighted sum equals zero. This is not as restrictive as it first seems because one can redefine the dynamics and disturbances to include the non-zero weighted sum as a constant bias (i.e. affine dynamics). }
\newline

[Assumption 2] The multistage, infinite horizon problem LQR problem, denoted $\mathbb{P}^{MS}_{LQR}$ is defined to have the following properties:
\begin{itemize}
    \item $\mathbb{P}^{MS}_{LQR}$ has the same robust horizon as in $\mathbb{P}^{MS}_N$
    \item the control horizon is extended to infinity
    \item the dynamics, stage and terminal costs, of $\mathbb{P}^{MS}_{LQR}$ and $\mathbb{P}^{MS}_N$ are the same.
    \item the additive disturbance is set to zero for the entire horizon
    \item the state and input constraints are dropped, i.e. $\mathcal{X}_k=R^{n_x}$ and  $\mathcal{U}_k=R^{n_u}$
\end{itemize}

With the assumptions above, it can be shown that the value function $\mathcal{V}^{MS}_{LQR}(x)$ is given by: 
\begin{align*}
     \mathcal{V}^{MS}_{LQR}(x) &= x^TP^{MS}_{LQR}x
\end{align*}
where $P^{MS}_{LQR}$ is the weighting matrix defined by the corresponding Riccati equation.
\newline
This allows us to present the first result of this paper: 
}

\begin{theorem}\label{thm-lqr-under}

Let the terminal cost of  $\mathbb{P}^{MS}_N$ be chosen such that \begin{align*}
   & x^TQ_fx \ge \mathcal{V}^{MS}_{LQR}(x)\qquad   \\
    & i.e.\ Q_f - P^{MS}_{LQR}\psd 0
\end{align*}
and let $d_{s,k}$ be chosen such that $\sum_s^S w_sd_{s,k}=0$.

Then $\mathcal{V}^{MS}_{LQR}(x)$ is an under-estimator of $\mathcal{V}^{MS}_N(x)$.
\end{theorem}
\begin{pf} 
The value function, $\mathcal{V}^{MS}_N$, can be explicitly written as a function of the scenario value function and disturbance:
\begin{equation*}
    \mathcal{V}^{MS}_N(\hat{x}) = \sum_{s=1}^S w_s \mathcal{V}_{N,s}(\hat{x}) = \sum_{s=1}^S w_s \mathcal{V}_{N}(\hat{x},d_{s})
\end{equation*}
where $d_{s}$ is a vector with elements $d_{k,s}$. As the value function is convex in $d_s$, by (the generalised) Jensen's inequality \citep{boyd2004book}:
\begin{equation*}
    \mathcal{V}_N(\hat{x},0)=\mathcal{V}_{N}\left(\hat{x},\sum_{s=1}^S w_sd_{s}\right) \le \sum_{s=1}^S w_s\mathcal{V}_{N}(\hat{x},d_{s}) \label{cnv-cost-eqnJensen}
\end{equation*}
where by assumption  $\sum_s^S w_sd_{s,k}=0$.

Note that if $Q_f=P^{MS}_{LQR}$ then $\mathcal{V}^{MS}_{LQR}\le \mathcal{V}_N(\hat{x},0)$ as $\mathbb{P}^{MS}_{LQR}$ is a relaxation -- it has the same objective, dynamics and data without state or input constraints.
 
For any other choice of  $Q_f$, by assumption  $x^TQ_fx > x^TP^{MS}_{LQR}x$ and hence $\mathcal{V}^{MS}_{LQR}$ remains an under-estimator.
    \qed

\end{pf}
Thus, by considering the associated LQR problem, a matrix $P$ can be found for use in \eqref{cnv-cost-eqns-icnn} that is a lower bound of $\mathcal{V}^{MS}_N$. 
Although considering the LQR problem without additive disturbances gives a lower bound, we note that under additional assumptions of the origin this choice exactly describes the curvature around the origin.

\begin{theorem}\label{thm- cons-bias}
Let $\mathbb{P}^{MS}_N$, $\mathcal{V}^{MS}_N$, $\mathcal{V}_N$, $\mathcal{V}^{MS}_{LQR}$ be defined as in Assumptions 1 and 2. Let the solution of $\mathbb{P}^{MS}_N$ in a neighbourhood $\mathcal{N}$ around the origin have no active inequality constraints. Then:
\begin{equation*}
   \mathcal{V}^{MS}_N(x) = \mathcal{V}^{MS}_{LQR}(x) + C, \qquad x \in \mathcal{N}
\end{equation*}
where $C\ge0$ is some scalar constant.
\end{theorem}
\begin{pf}
     As $\mathbb{P}^{MS}_N$ has no active inequality constraints, 
    \begin{equation*}
        \mathcal{V}^{MS}_{LQR}(x) = \mathcal{V}_N(x,0),\  x \in \mathcal{N}
    \end{equation*}
    Furthermore, as the active constraints do not change in $\mathcal{N}$, $u$ is a linear function of $x$, i.e. $u=Kx\ \forall x\in\mathcal{N}$. For all $s\in S$, let $\Tilde{A}_{k,s}=A_{k,s}+B_{k,s}K$.
    Then due to linearity, the state evolution is given by:
    \begin{align*}
    x_{k+1,s} = \Tilde{A}_{k,s}\dots\Tilde{A}_{0,s}x_{0,s} + d_{k,s} +\sum_{t=0}^{k-2}\Tilde{A}_{k-1,s}\dots\Tilde{A}_{t+1,s}d_{t,s}\:\: \forall x_{0,s}\in\mathcal{N}, s=1,\dots,S
    \end{align*}
    Note that if $x_{0,s}=0$ the dynamics are solely due to the sequence $d_{1:N,s}$. Accordingly the value function can be decomposed as:
     \begin{equation*}
        \mathcal{V}^{MS}_N(\hat{x})  = \mathcal{V}_N(\hat{x},0) +  \mathcal{V}^{MS}_N(0),\qquad
        \hat{x} \in \mathcal{N}
    \end{equation*}    \qed
\end{pf}
{\color{black}
When approximating the value function, it is a major concern that the controller will show good performance near the origin. Theorem \ref{thm- cons-bias} shows that if the surrogate is constructed as proposed, then as long as $\hat{\mathcal{V}}_{sur}$ learns a constant bias around the origin, there will be zero error in the controller output in a neighbourhood around the origin. Although the assumption of no active inequality constraints at the origin is restrictive, it is not an uncommon assumption in the MPC literature, e.g. \citep{limon2006stability, muske1993model}.  
}

\subsection{Learning feasibility}
{\color{black}In the section above, a tacit assumption is that any $x_1$ in the state constraint set $\mathcal{X}_1$, see \eqref{eq:  constraint sets}, is feasible for \eqref{cnv-cost-eqnms-mpc}. If so then one can simply sample points in  $\mathcal{X}_1$ to train the network. This is not necessarily true due to the system dynamics and other inequality constraints.} To address this, we propose to find a separate convex approximator to learn a penalty term that describes the feasible region. 
This will ensure feasibility. 

To learn feasibility we assume that a soft-constrained problem is used to generate the training data. This makes it possible to generate points that are outside the feasible space of the original MPC problem, such that the behaviour for these regions also can be learned and penalized in our one-step approach. 
For example, a nominal MPC problem with box constraints on $x$, can be reformulated as: 
\begin{subequations}\label{cnv-cost-eqnpenal-mpc}
\begin{align}
   \mathcal{V}_{N,\mu}(\hat{x},\mu) &= \min_{u,x}\sum_{k=0}^{N-1} l_k(x_k,u_k)+V_t(x_N)\nonumber\\
    &+\mu\sum_{k=1}^{N} \|\eta^u_{k}+\eta^l_{k}\|_1\\
    x_0 &= \hat{x}\\
    x_{k+1} &= Ax_k+Bu_k,\:\: k=0,\dots,N-1\label{cnv-cost-eqnnom-dynamics}\\
    u_k &\in \mathcal{U}_k,\qquad k=0,\dots,N-1\\
    x_k &\le x^{u} + \eta^u_k, \qquad k=1,\dots,N \\
   x_k &\ge  x^{l} - \eta^l_k , \qquad k=1,\dots,N \\
   0&\le\eta^u_k,\qquad 0\le\eta^u_l
\end{align}
\end{subequations}
where $\eta^u_k$ and $\eta^l_k$ are slack variables,  $\mu$ is a penalty parameter and $\mathcal{V}_{N,\mu}(\hat{x},\mu)$ is the optimal value function. As an exact penalty is used the optimum of \eqref{cnv-cost-eqnpenal-mpc} is equivalent to that of \eqref{cnv-cost-eqnstandard-mpc} if  $\mu>\mu^*$ where $\mu^*$ is the largest Lagrange multiplier arising from the bound constraints of  \eqref{cnv-cost-eqnstandard-mpc} \citep{Nocedal2006}.

The cost-to-go, $\mathcal{V}_{N-1,\mu}$ may be decomposed as:
\begin{equation}
    \mathcal{V}_{N-1,\mu}(\hat{x},\mu) =\mathcal{V}_{N-1}(\hat{x}) + \mu\mathcal{F}_{N-1}(\hat{x})
\end{equation}
where $\mu\mathcal{F}_{N}$ is the convex, piecewise linear contribution of the slack variables to the value function. 

Moving forward, there are two possibilities for learning feasibility. One can either develop a surrogate to describe the whole function $\mathcal{V}_{N-1,\mu}$ or two separate surrogates for the decomposed problem. 

In this work, we consider the use of two surrogates, because our requirements of accuracy of the two terms are different.  For example, the approximation of $\mathcal{V}_{N-1}$ does not have to be accurate in the infeasible region, where  $\mathcal{F}_{N-1}$ is non-zero. Furthermore, the approximation of $\mathcal{F}_{N-1}$ only needs to be accurate near the boundary of the feasible region. Any inaccuracy in the interior of the infeasible region can be captured by using a larger  $\mu$ in the one-step problem \eqref{cnv-cost-eqn1stage-mpc} (this term can be adjusted after optimisation of the surrogates).

\section{Fitting a surrogate}\label{cnv-cost-sec: approximating=cnvx-fns}
We consider the task of fitting a convex surrogate to data generated by an unknown convex function. {In doing so we follow the approach outline in Algorithm \ref{alg: prop}.   \color{black} We assume that the cost-to-go can be decomposed into a multistage LQR term and convex non-linear term. To simplify the notation, we assume that surrogates of the value function and feasibility term are structurally the same but have different parameters. Once the surrogates have been fitted, we form the 1-step ahead multistage MPC problem:

\begin{subequations}\label{cnv-cost-eq: cnv-surr-mpc-reform}
    \begin{align}
    & \min_{u_0,x,{V},F}\:\:  l_0(\hat{x},u_0)+ \sum_{s=1}^{S} w_s\left({V}_{s} + x_{1,s}^TP^{MS}_{LQR}x_{1,s}\right)+\mu{F}_{s}\\
    &x_{1,s} = A_{0,s}\hat{x}+B_{0,s}u_{0} + d_{0,s},\qquad  s=1 \dots S\\
    &{V}_{s} \ge f(x_s,\theta_\mathcal{V}),\qquad   s=1 \dots n_s \label{eq: v-surrogate-here}\\
    &{F}_{s} \ge f(x_s,\theta_\mathcal{F}),\qquad   s=1 \dots n_s \label{eq: f-surrogate-here}\\
    & u_0 \in \mathcal{U}_0,\:\: x_{1,s} \in \mathcal{X}_1,\:\: 0 \le {V}_{s},\:\: 0 \le {F}_{s},\   s=1 \dots n_s
\end{align}
\end{subequations}
where $V_s$ and $F_s$ are new variables that approximate the functions $\mathcal{V}_{N-1}$ and $\mathcal{F}_{N-1}$ evaluated at $x_{1,s}$ due to constraints \eqref{eq: v-surrogate-here} and \eqref{eq: f-surrogate-here}. Although these are inequality constraints, these constraints will be active as $V_s$ and $F_s$ are positive contributions in the objective. 
Note that the same surrogate is evaluated for each scenario $s$. 

In the following section, we present two approaches for fitting a convex surrogate (line 7, Algorithm \ref{alg: prop}) -- namely, (1) fitting an interpolating convex function and (2) fitting an input convex neural network. The same approachs can be used for fitting any convex data, and so we, for convenience, present the approaches only for fitting the cost-to-go data. We use a least squares objective when fitting the surrogates, but this is not a prescriptive choice.

\begin{algorithm}
\caption{Proposed approach}\label{alg: prop}
\begin{algorithmic}[1]
\Require $\texttt{mpc}_{N-1}(\hat{x})$  \Comment{MPC solution (reduced horizon), given $\hat{x}$}
\Require $\hat{\mathcal{X}}$ \Comment{Set of initial conditions}
\Require $f: \mathbb{R}^{n_x}\times \mathbb{R}^{n_\theta} \to\mathbb{R}$ \Comment{Choice of surrogate}
\State $Data \gets \emptyset$, $m \gets \texttt{length}(\hat{\mathcal{X}})$
\While {$m \neq 0$} \Comment{Generate the cost-to-go dataset}
\State $\mathcal{V},\mathcal{F} \gets \texttt{mpc}_{N-1}(\hat{\mathcal{X}}_m) $
\State $Data \gets Data \cup \{\hat{\mathcal{X}}_m,\mathcal{V},\mathcal{F}\}$
\State $m \gets m - 1$
\EndWhile
\State $\theta_\mathcal{V},\theta_\mathcal{F} \gets \texttt{fit}(f,Data)$
\end{algorithmic}
\end{algorithm}
\

}

\subsection{\color{black} Interpolating convex function}
{\color{black} The following is summarised from Chapter 6 of \citet{boyd2004book}. 
Consider an arbitrary convex function $f:\mathbb{R}^{n_x}\to\mathbb{R}$ that exactly interpolates some convex data:
\begin{align}\label{cnv-cost-eqn: infinite-dim-fit-convex}
    &f(x_i) = \mathcal{V}_i, \qquad i=1,\dots,m 
\end{align}
where $x_i\in\mathbb{R}^{n_x},\ \mathcal{V}_i=\mathcal{V}(x_i)$. From the definition of convexity, this is the case if and only if there exists vectors $g_1$,\dots,$g_m\in\mathbb{R}^{n_x}$ such that:
\begin{align}\label{eq: g-condition}
        f(x_j) \ge  f(x_i) + g_i^T(x_j-x_i),\qquad  i,j=1,\dots,m
\end{align}
Given $g_1$,\dots,$g_m$ satisfying \eqref{eq: g-condition}, one can construct various convex functions that perfectly interpolate the data. To find these vectors, we can solve:}
\begin{subequations}\label{cnv-cost-eqn: finite-dim-fit-convex}
\begin{align}
    &\min_{\hat{\mathcal{V}}_i g_i}\:\: \sum_{i=1}^m (\mathcal{V}_i - \hat{\mathcal{V}}_i)^2\\
    &\text{s.t. } \hat{\mathcal{V}}_j \ge \hat{\mathcal{V}}_i + g_i^T(x_j-x_i),\quad  i,j=1,\dots,m
\end{align}
\end{subequations}
where $\hat{\mathcal{V}}_i=f(x_i)$\footnote{Theoretically the optimal $\hat{\mathcal{V}}_i$ should equal $\mathcal{V}_i$, and hence the optimum objective value should be zero. However, for numerical reasons, there may be small errors, and hence the equality may not hold.}. Once \eqref{cnv-cost-eqn: finite-dim-fit-convex} is solved one can use the optimal values ($\hat{\mathcal{V}}_i^*$, $g_i^*$) to \textit{construct} arbitrary convex functions that interpolate the data, e.g. a piecewise affine function, $f_{pwa}$ is defined by:
\begin{subequations}\label{cnv-cost-eqnconvex-pwa-boyd}
\begin{align}
    f_{pwa}(\hat{x}) = &\min_V V\\
    &\text{s.t. } V \ge \hat{\mathcal{V}}^*_i + g_i^{*T}(x_i-\hat{x}),\:  i=1,\dots,m
\end{align}
\end{subequations}
\eqref{cnv-cost-eqnconvex-pwa-boyd} is a linear program with $m$ inequality constraints and one variable, $V$. {\color{black}Thus for the multistage MPC problem we can form the one step ahead problem \eqref{cnv-cost-eq: cnv-surr-mpc-reform}, with the following instead of $\eqref{eq: v-surrogate-here}$ (neglecting the feasibility surrogate):
\begin{align}\label{cnv-cost-eq: interp-mpc-reform}
    &{V}_{s} \ge  \hat{\mathcal{V}}^*_i + g_{\mathcal{V},i}^{*T}(x_{1,s} - x_{i}),\   s=1 \dots S,\ i=1,\dots,\bar{m}
\end{align}
where $\theta = [\hat{\mathcal{V}}^*_i,\: g_{\mathcal{V},i} ]$}.
\eqref{cnv-cost-eq: interp-mpc-reform}  introduces $n_s{m}$ new inequality constraints. 
Thus, although this approach avoids forming the full scenario tree, it still involves including many new inequality constraints, which can increase the computational cost. {\color{black}Because of this, we expect that this approach will have to be combined with some heuristic to select only $m^*<<m$ constraints, e.g. by neglecting constraints corresponding to points that are far away.} 

\subsection{Input-convex neural networks}
 \begin{figure}
     \centering
     \includegraphics[width=\linewidth]{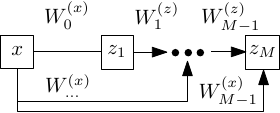}
     \caption{Schematic of a feedforward input-convex neural network of $M$ layers. For simplicity $z_0$, $W^{(z)}_{0}$, and bias blocks are not shown.}
     \label{cnv-cost-fig: icnn}
 \end{figure}
Practically is reasonable to consider finding an approximate surrogate instead of an exact interpolating function. As such we consider training input-convex neural networks (ICNNs). \citep{Amos2017,seel2022convex, yang-icnn}.  

For ease of exposition, we consider a $M$-layer, fully connected ICNN (\figref{cnv-cost-fig: icnn}). This network, $\nn$, 
is defined as:
\begin{subequations}\label{cnv-cost-eqn: nn}
\begin{align}
         &\nn(x,\theta)= z_M\\
        & z_{i+1}= \alpha_i(W^{(z)}_iz_i + W^{(x)}_ix + b_i), \: i=0,\dots, M-1 \\
         &0\le W^{(z)}_{i}, \qquad i=1,\dots,M-1 \label{eq-convex-weight-req}\\
         &\alpha_{i}\text{ convex and non-decreasing}, \quad i=0,\dots,M-1 
\end{align}
\end{subequations}
where $z_i$ are the activations of layer $i$, $\alpha_i$ is that layer's activation function, and $\theta=\{W^{(z)}_{0:M-1},\ W^{(x)}_{0:M-1},\ b_{0:M-1}\}$ are the parameters, and that $z_0\equiv0,\ W^{(z)}_{0} \equiv 0$.

As the elements of $W^{(z)}_{1:M-1}$ are non-negative, and $\alpha_{0:M-1}$ are convex and non-decreasing,  $\nn$ is convex with respect to $x$ \citep{Amos2017}. This is because: (1) the composition of a convex and convex non-decreasing function is convex,
and (2) non-negative weighted sums of convex functions preserve convexity \citep{boyd2004book}. The network $f_{NN}$ can be trained by any algorithm, as long as the constraint \eqref{eq-convex-weight-req} is satisfied, e.g. stochastic descent with projection. One can convert this constraint into a penalty in the objective, however one must ensure that the constraint is satisfied; otherwise, the neural network is not guaranteed to be convex. As ICNNs are convex, they are easy to optimize over, very data efficient compared to normal neural networks, and often don't require further regularization \citep{Amos2017, yang-icnn}. 
Similarly to \eqref{cnv-cost-eq: interp-mpc-reform}, after training we can form the one step ahead problem
{\color{black}\eqref{cnv-cost-eq: cnv-surr-mpc-reform}, with the following instead of $\eqref{eq: v-surrogate-here}$ (neglecting the feasibility surrogate):}
    \begin{align}\label{cnv-cost-eq: icnn-mpc-reform}
    &{V}_{s} \ge f_{NN}(x_{1,s},\theta),\qquad   s=1 \dots n_s
\end{align}
{where $\theta$ are now parameters, and not variables. Note that one can either include the neural network equations in the one step ahead problem or include the neural network as a convex non-linear function.}

Lastly, we note that if one considers $M=2$, $\alpha_{1}(y)=y$, $W^{(z)}_1=I$, and  $\alpha_{2}(y)=\max\{y_{i},\dots,y_{n_z}\}$, where $n_z$ is the hidden layer width, (i.e. a max-out layer) then the ICNN describes an alternative parameterisation of any convex piecewise-affine (PWA) function defined by \eqref{cnv-cost-eqnconvex-pwa-boyd}.

\section{Numerical results}\label{cnv-cost-sec: results}
To demonstrate our proposed approaches we consider {\color{black}two demonstrative case studies}. The code has been implemented in Julia, and primarily makes significant use of the algebraic modelling language \texttt{JuMP} \citep{JuMP-paper},  and the solvers \texttt{IPOPT} \citep{ipopt-Wachter2006} and \texttt{L-BFGS} \citep{LBFGS-Liu1989}. 

 {\color{black}
 For both case studies we use \texttt{IPOPT} to solve the full and 1-step horizon MPC problems. To allow for sampling from infeasible initial conditions and to generate the feasibility surrogate we reformulate the state inequalities as soft constraints and positively constrained slack variables. These variables are included in the objective with a weighting of $10^4$. We use the same model in the MPC formulations and for simulating the system -- i.e. there is no model mismatch.
 
\subsection{Case study 1: QCQP MPC}
In this case study, we demonstrate the proposed methods on a (convex) quadratically constrained quadratic program (QCQP). We compare the proposed formulation against solving the full problem, and solving the 1-step problem with only the LQR cost-to-go.  To easily visualize the closed loop trajectories we consider a two-state example, namely the control of a chemostat with substrate inhibition and linearized dynamics \citep{pappas2021multiparametric}:
\begin{subequations}
\begin{align}
    A_k = A &=  \begin{bmatrix}
 0.9875 & 0.1601  \\
-0.1327 & 0.7743 
 \end{bmatrix}\\
 B_k = B &= \begin{bmatrix}
     -0.2409\\ -0.6980
 \end{bmatrix}
\end{align}
\end{subequations}
Following \citet{pappas2021multiparametric}, we formulate the control problem as a QCQP, with the goal to regulate the state to the origin (the linearization point). We consider uncertainty by fully branching the scenario tree and minimizing the average cost of all scenario, yielding the full-horizon problem:
\begin{subequations}
\begin{align}
\min_{u,x}&  \sum_{s=1}^{S}w_s\mathcal{V}_{N,s}\\
& \mathcal{V}_{N,s} = \sum_{k=0}^{N-1}(x_{k,s}^TQx_{k,s} + u_{k,s}^TRu_{k,s}) + x_{N,s}^TPx_{N,s}\:\: s=1,\dots,S \\
&x_{k+1,s}= Ax_{k,s} +Bu_{k,s}+ d_{k,s},\qquad k=0,\dots,N-1,\:\: s=1,\dots,S \\
& \begin{bmatrix}
     -1.5302\\0.1746
 \end{bmatrix} \le x_{k,s} \le \begin{bmatrix}
     0.4698\\ 1.8254
 \end{bmatrix}\qquad k=0,\dots,N,\:\:s=1,\dots,S\\
 &-0.3 \le u_{k,s} \le 0.7,\qquad k=0,\dots,N-1,\:\:s=1,\dots,S\\
 &x_{0,s} = \hat{x} \qquad s=1,\dots,S\\
 &\begin{bmatrix}
     -0.0050\\ -0.0531
 \end{bmatrix}\le d_{k,s} \le \begin{bmatrix}
     0.0050\\ 0.0531
 \end{bmatrix}\qquad k=0,\dots,N-1,\:\:s=1,\dots,S\\
 &\sum_{k=0}^{N-1}u_{k,s}^2 \le 0.2, \qquad k=0,\dots,N-1,\:\: s=1,\dots,S\\
 & x_{N,s}^T\begin{bmatrix}
     1 & 0\\ 0 & 12.25
 \end{bmatrix}x_{N,s} + \begin{bmatrix}
     1 & -2.1
 \end{bmatrix}x_{N,s} \le 0.66\:\:s=1,\dots,S\\
    & x_{k,s_1} = x_{k,s_2} \Rightarrow   u_{k,s_1} = u_{k,s_2},\:\:   k=0,\dots,N-1, \: s_1, s_2=1,\dots,S 
\end{align}
\end{subequations}
with the objective function weightings $w_s=1/N$, $Q=1000I$, $R=0.01$, and $P=\begin{bmatrix}
     7 812.7 & 3 091.4\\ 3 091.4 & 2 402.8
 \end{bmatrix}$ (the LQR cost-to-go of the nominal problem). We consider a control and robust horizon of 5. The scenario tree consists of the vertex values of $d_k$, and defines $(2^5)^2=1024$ scenarios. 

\subsubsection{Training the surrogate}
{\emph{Data generation}}\\
To generate data to fit the surrogates, we need to solve the control problem with a horizon of $N-1$ to generate training data (line 3, Algorithm \ref{alg: prop}). For training we generate a $20\times 20$ equidistant grid of initial conditions with the ranges $-1.7802\le x_1 \le 0.7198$ and $-0.4246 \le x_2 \le 2.0754$. For the size of the problem this is a dense grid, and the surrogates' approximation error is expected to be small.  The input and output data of the surrogates are normalised to have a range between zero and one. We use the same procedure to generate test data, but solve the problem with the full horizon. 

{\emph{Fitting the surrogates}}\\
To fit the interpolating convex functions we solve \eqref{cnv-cost-eqn: finite-dim-fit-convex}. Although theoretically the optimal objective should be zero, due to numerical reasons the optimizer cannot perfectly fit the data. 

For the neural networks approximating the value and feasibility functions we use a single hidden layer of width 20 (i.e. $M=2$), with activation functions:
\begin{equation}
             \alpha_1(x) = \max(0.01x,x), \qquad \alpha_2(x) = \max(0.0,x)
 \end{equation}
We select this choice of $\alpha_2$ as the network output is non-negative.
When training the neural networks we use an objective function made up of the sum of squared errors plus the absolute value of the neural network evaluated at the origin, e.g. for the value function approximation:
\begin{equation}
    \min_\theta \sum_{i=1}^m \left(f_{NN} (x_i,\theta) - \mathcal{V}^{sur}_{N-1}(x_i)\right) + |f_{NN}(0,\theta)|
\end{equation}
To train the ICNNs we use  L-BFGS \citep{LBFGS-Liu1989}, and enforce the convexity requirement \eqref{eq-convex-weight-req} by specifying lower bounds on the network weights. 

\subsubsection{Numerical results}

\emph{Approximation error and solution time}\\
The approximation error of using the ICNN and interpolating formulation in \eqref{cnv-cost-eq: cnv-surr-mpc-reform} is shown in is shown in Figure \ref{fig: error-in-u-qcqp}. The interpolating formulation has marginally smaller errors than the ICNN, although both formulations have a tail of relatively larger errors in $u$. These errors are primarily due to mispredictions near the boundary of the feasible space. Note that $\mathcal{F}$ is piecewise linear, with a value of zero within the feasible region. Thus, the ``kink" defining the feasible region will only be exactly predicted if it is a data point in the training data. 

Examining the solve times, both surrogates result in a significant reduction in computational time -- the interpolating formulation is able to achieve an order of magnitude reduction, while the ICNN improves by three orders of magnitude. (Figure \ref{fig: solve-times-qcqp}). As discussed earlier, the poorer speed up of the interpolating formulation is likely due to the introduction of the additional inequalities.  

\begin{figure}
    \centering
    \includegraphics[width=.8\textwidth]{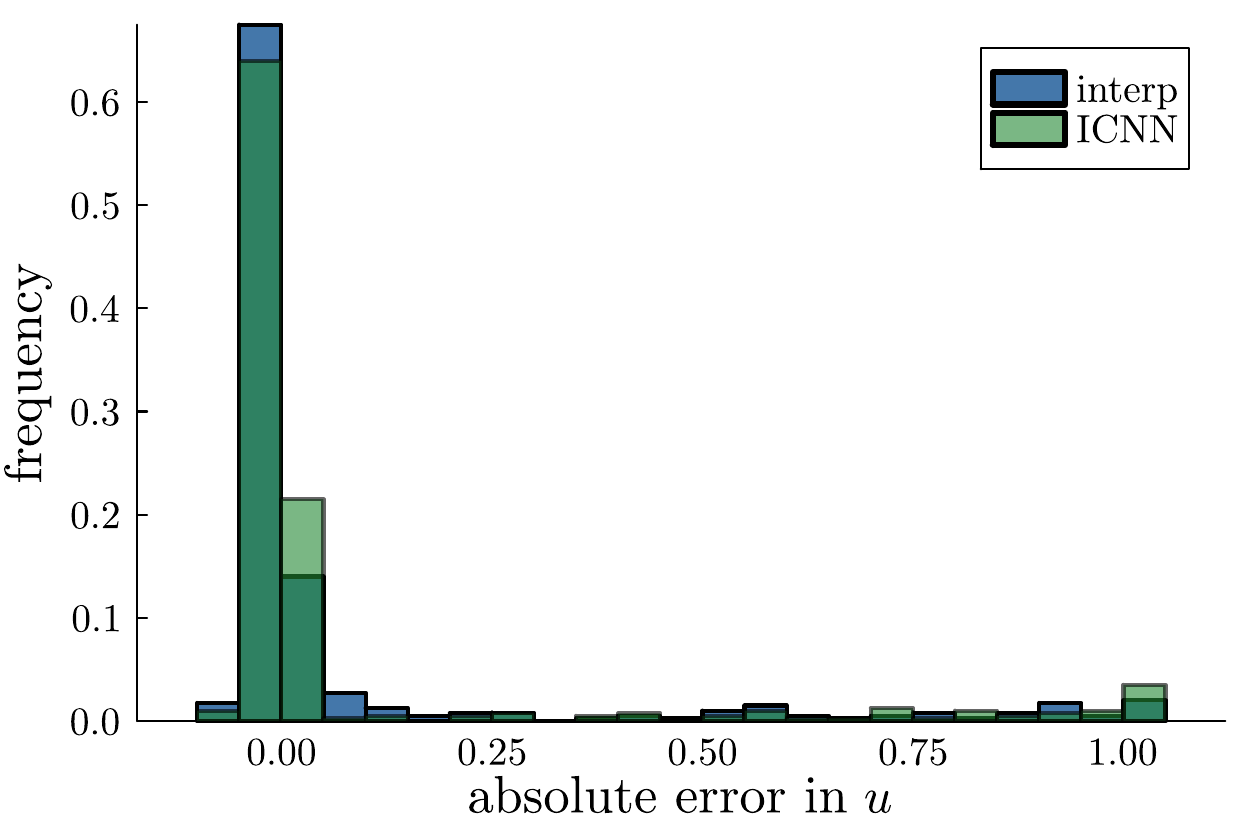}
    \caption{Error in the approximation of $u$ using ICNNs and interpolating convex functions.}
    \label{fig: error-in-u-qcqp}
\end{figure}

\begin{figure}
    \centering
    \includegraphics[width=.8\textwidth]{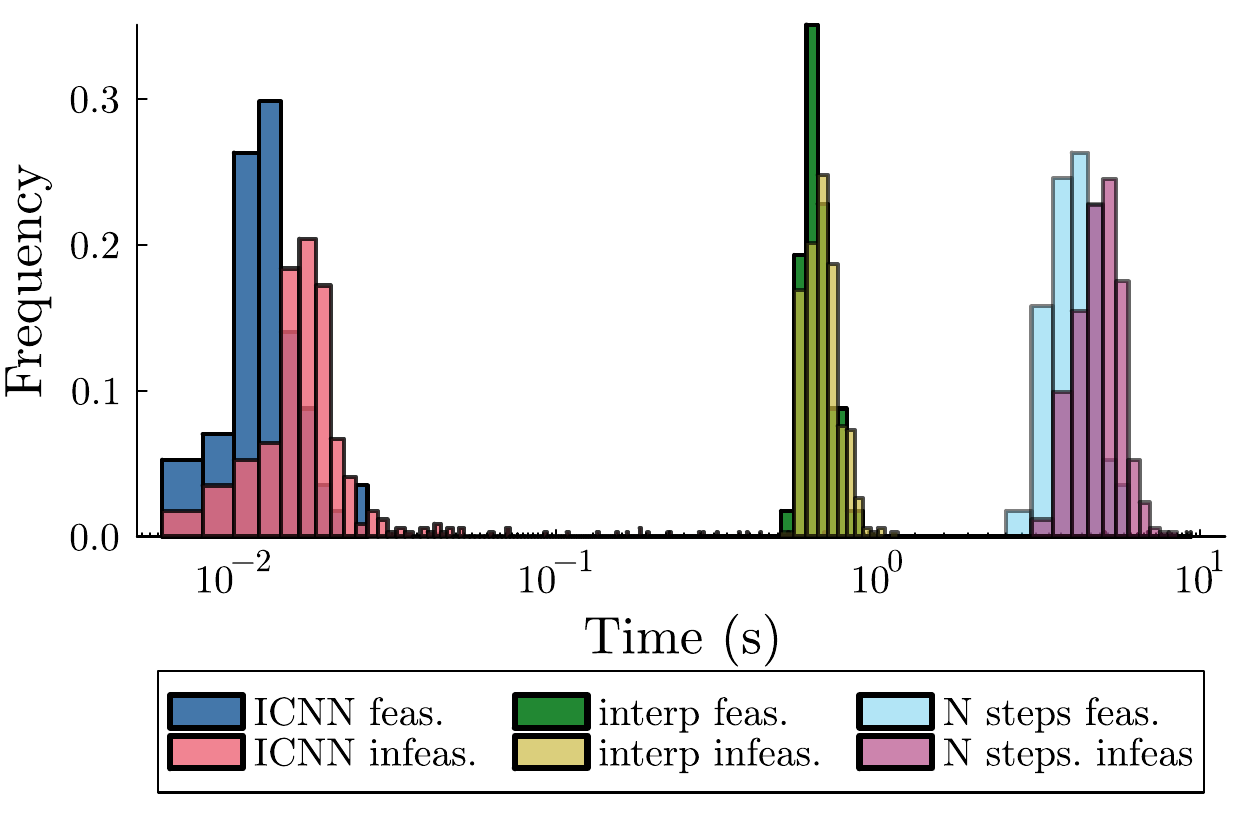}

    \caption{Solve times.}
    \label{fig: solve-times-qcqp}
\end{figure}
\emph{Closed loop performance}\\
We show two examples of system trajectories to demonstrate the closed-loop performance and contrast it to using a 1-step MPC formulation with only the LQR cost-to-go. As the two surrogates have a similar distribution of errors (Figure \ref{fig: error-in-u-qcqp}), we show the closed-loop performance of only the ICNN. 
We illustrate the control performances with example trajectories related to two different starting points. 
The first example demonstrates the effect of active constraints, and the second example considers a starting point that remains unconstrained.

In the first example, Figure \ref{fig: closed-loop-comparison-LQR-prop-1}, the system starts from a feasible point but has a long period with active constraints. The MPC with ICNN cost-to-go briefly violates the state constraint at one step but successfully guides the system towards the origin. The brief constraint violation is caused by the approximation error observed near the boundary of the feasible region. In contrast, using only the LQR cost-to-go the controller keeps the system at the upper bound for a longer period, and eventually loses control and fails to drive the system to the origin within 60 steps.
\begin{figure}[h]
\centering
    \includegraphics[width=.8\textwidth]{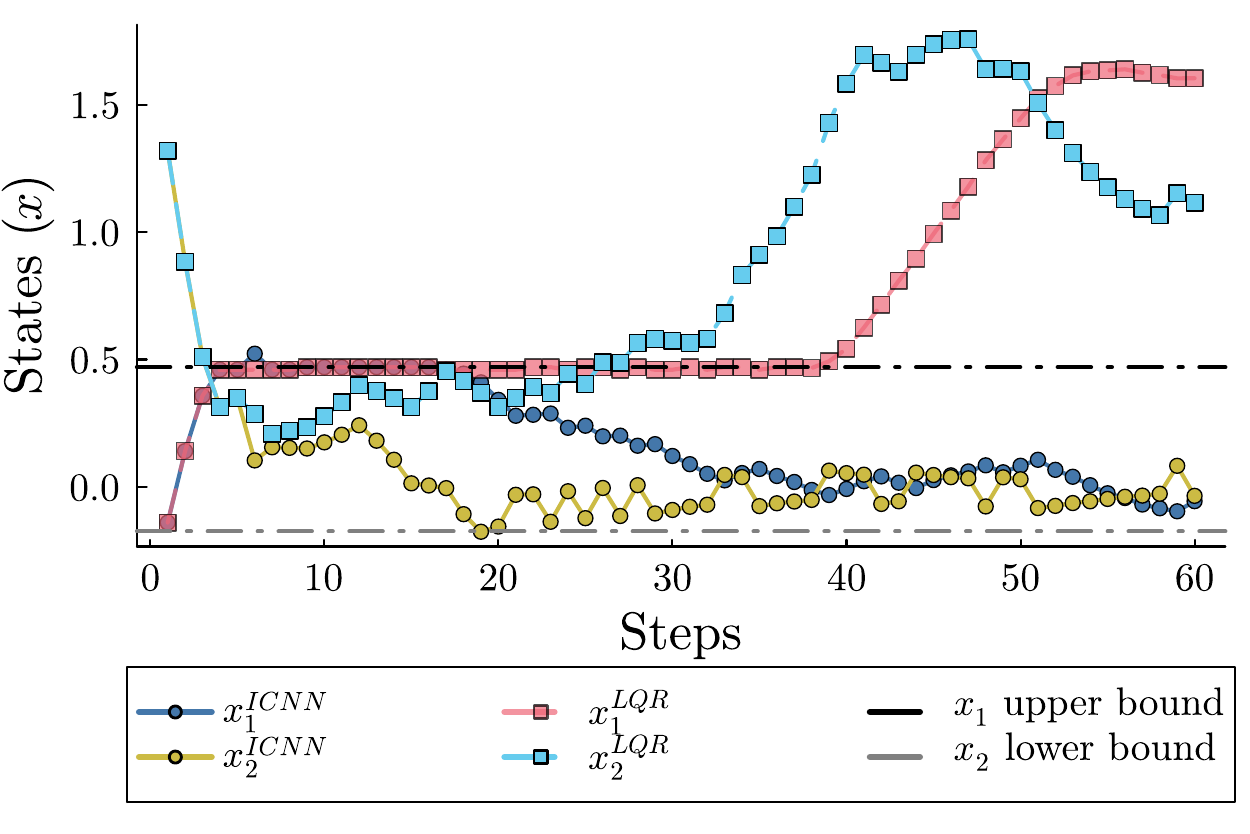}
    \caption{Comparison of closed-loop trajectories of the 1-step MPC with initial state far from the origin and near the bounds.}
    \label{fig: closed-loop-comparison-LQR-prop-1}
\end{figure}

In contrast in the second example (Figure \ref{fig: closed-loop-comparison-LQR-prop-2}) the system starts and is kept near the origin, and both MPC formulations have exactly the same trajectories. This behaviour is expected as in the area around the origin there are no active constraints, and so Theorem 2 applies -- i.e. the LQR cost-to-go exactly approximates the full horizon, and so the surrogate learns to apply no additional correction.
\begin{figure}[h]
\centering
    \includegraphics[width=.8\textwidth]{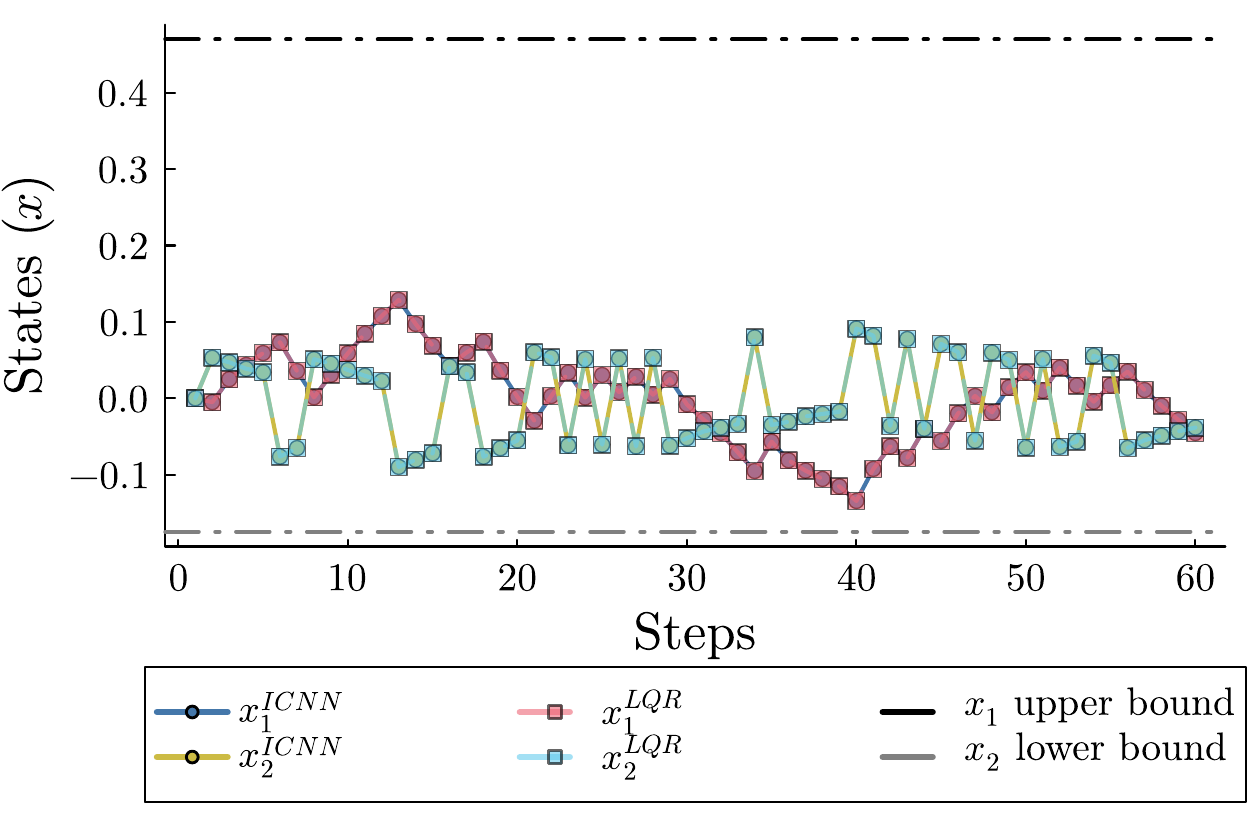}
    \caption{Demonstration that with an initial state near the origin, and appropriate assumptions, the closed-loop trajectories of the 1-step MPC with surrogate term and 1-step MPC with only the LQR are identical.}
    \label{fig: closed-loop-comparison-LQR-prop-2}
\end{figure}

}

\subsection{Case study 2}
{\color{black}The motivation of the second case study is primarily to examine the data-efficiency of the surrogates, and for initial results in exploring the influence of problem size. For this, we consider a four-state linear MPC problem. Although the problem is only a little larger, it is already large enough to show the expected trends in data efficiency and problem size. 

We} consider a linearized model of a continuous stirred-tank reactor (CSTR)  with the reactions:
\begin{align*}
    &{A \rightarrow B \rightarrow C} \\
  &{2A \rightarrow D}
\end{align*}
The linearized model is described in  \citep{subramanian2021tube} and is in the form of \eqref{cnv-cost-eqnstandard-mpc nom-dynamics}, with $x=[\Delta C_a\ \Delta C_b\ \Delta T_R\ \Delta T_J]^T$, $u=\Delta F$, and 
\begin{align*}
    A_k = A &=  \begin{bmatrix}
 0.4 & -0.09 & -0.01 & 0. \\
 0.2 & 0.39 & 0.002 & 0. \\
 0.33 & 0.26 & 1.10 & 0.15 \\
 0.05 & 0.07 & 0.13 & 0.68
 \end{bmatrix}\\
 B_k = B &= \begin{bmatrix}
     0.1\\ -0.05\\ 0.8 \\ 0.1
 \end{bmatrix}\\
&\|d_k\|_\infty \le 0.1
\end{align*}
The state and input constraints are $[-5\ -5\ -3\ -5]\le x$ $\le[5\ 5\ 3\ 5]$, and $|u| \le 2$. As the objective we chose $Q=I$, $R=0.01I$, and $P$ as the LQR cost-to-go of the nominal problem. We consider a control horizon of $N=6$,  a robust horizon of $N_R=2$, and consider the vertex values of $d_k$ {\color{black} in the scenario tree.}. This corresponds to a scenario tree of $(2^4)^2=256$ scenarios. 

We consider approximating the value function contribution, $\mathcal{V}^{sur}_{N-1}$ and feasibility function $\mathcal{F}_{N-1}$, using {\color{black} the interpolating formulation, ICNN, and standard neural networks}. We also consider training a neural network in an imitation learning framework to learn the policy approximation $u=f_{NN}(x,\theta)$. While the former approach requires solving an optimisation problem to evaluate the control action, the latter only requires evaluating the trained neural network. {\color{black}However, as briefly discussed earlier by solving the 1-step ahead problem we provide structure to the approximation and so expect the latter approach to (in-general) require more data \citep{recht2019tour}.}

\subsubsection{Training the surrogates}

\emph{Data generation}

{\color{black}Data generation follows the same strategy as in Section 5.1. We solve the control problem with a control horizon of $N-1=5$, $N_R-1=1$, on a $5\times 5 \times 5 \times 5$ equidistant grid to generate 625 training points using \texttt{IPOPT} \citep{ipopt-Wachter2006}.
As before, we} allow for sampling from infeasible start points we reformulate the state inequality as soft constraints with positively constrained slack variables. The slack variables are included in the objective with a weighting of $10^4$. To evaluate our surrogates' performance, we solve the full problem ($N=6$, $N_R=2$) at the same grid points. The same grid points are used for training all of the surrogates, and we normalise the data to have a range between zero and one.
{\color{black}\newline\emph{Fitting the surrogates}\\
As before, to fit the interpolating surrogate we solve \eqref{cnv-cost-eqn: finite-dim-fit-convex}. We use the same neural network architecture and objective as in 5.1. Both the ICNN and standard neural networks are trained on the same data with L-BFGS, with the convexity constraint only applied to the ICNN.
}

We also train a neural network to approximate the policy function implicitly defined by the MPC problem. We use the same grid points but now train the network based on the least squares error in its approximation of $u$, i.e. with the objective
\begin{equation}
    \min_\theta \sum_{i=1}^m (f_{NN}(x_i,\theta)-u^*(x_i))^2
\end{equation}
For training we use NADAM with 20 000 iterations and a mini-batch size of 50. For consistency with the other neural network surrogates, we use the same architecture apart from the activation functions. For the policy approximation we use $\alpha_1(x)=\tanh(x)$, and $\alpha_2(x)=2\tanh(x)$. This choice means that the network satisfies the constraint on the control output by design. 
\subsection{Performance of the surrogates}

\subsubsection{ICNN and interpolating convex function}
We first compare the implementation and performance of the ICNNs and interpolating convex functions. 
For both surrogates, we formulate the 1-step ahead MPC problem \eqref{cnv-cost-eq: interp-mpc-reform} and \eqref{cnv-cost-eq: icnn-mpc-reform} in JuMP \citep{JuMP-paper} and use the solver \texttt{IPOPT} \citep{ipopt-Wachter2006}. For stable performance of the solver, we use the Mehrotra algorithm option when solving \eqref{cnv-cost-eq: interp-mpc-reform} and use the limited memory Hessian approximation option when solving \eqref{cnv-cost-eq: icnn-mpc-reform}.

\begin{figure}
    \centering
        \includegraphics[width=.9\textwidth]{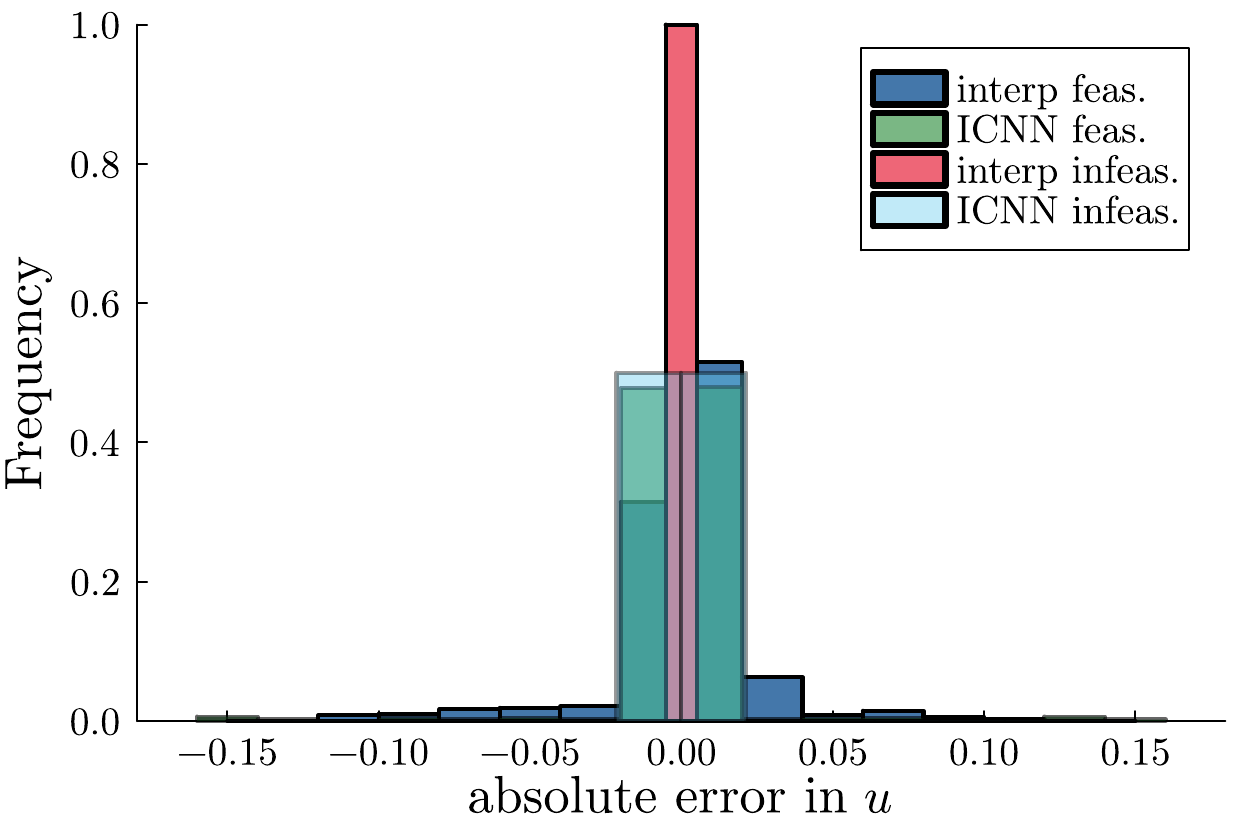}
    \caption{Error in the approximation of $u$ using ICNNs and interpolating convex functions, from feasible and infeasible start points.}
    \label{fig: error-in-u-icnn-interp}
\end{figure}

The approximation error in the control output when using the ICNNs and interpolating convex functions is shown in Figure \ref{fig: error-in-u-icnn-interp}. Figure \ref{fig: error-in-u-icnn-interp} shows that from infeasible start points the approximation error is below the tolerance used to generate the solutions. This is because in the infeasible region, the optimal action is to saturate the controller to leave the infeasible region. From feasible starting points the ICNN has a very good average approximation error, although there are some significant outliers of $\pm 0.15$ (Figure \ref{fig: error-in-u-icnn-interp}). In contrast, the interpolating function results in small, mostly positive errors in the control output.

\begin{figure}
    \centering

    \begin{subfigure}{0.95\columnwidth}
    \includegraphics[width=1.\linewidth]{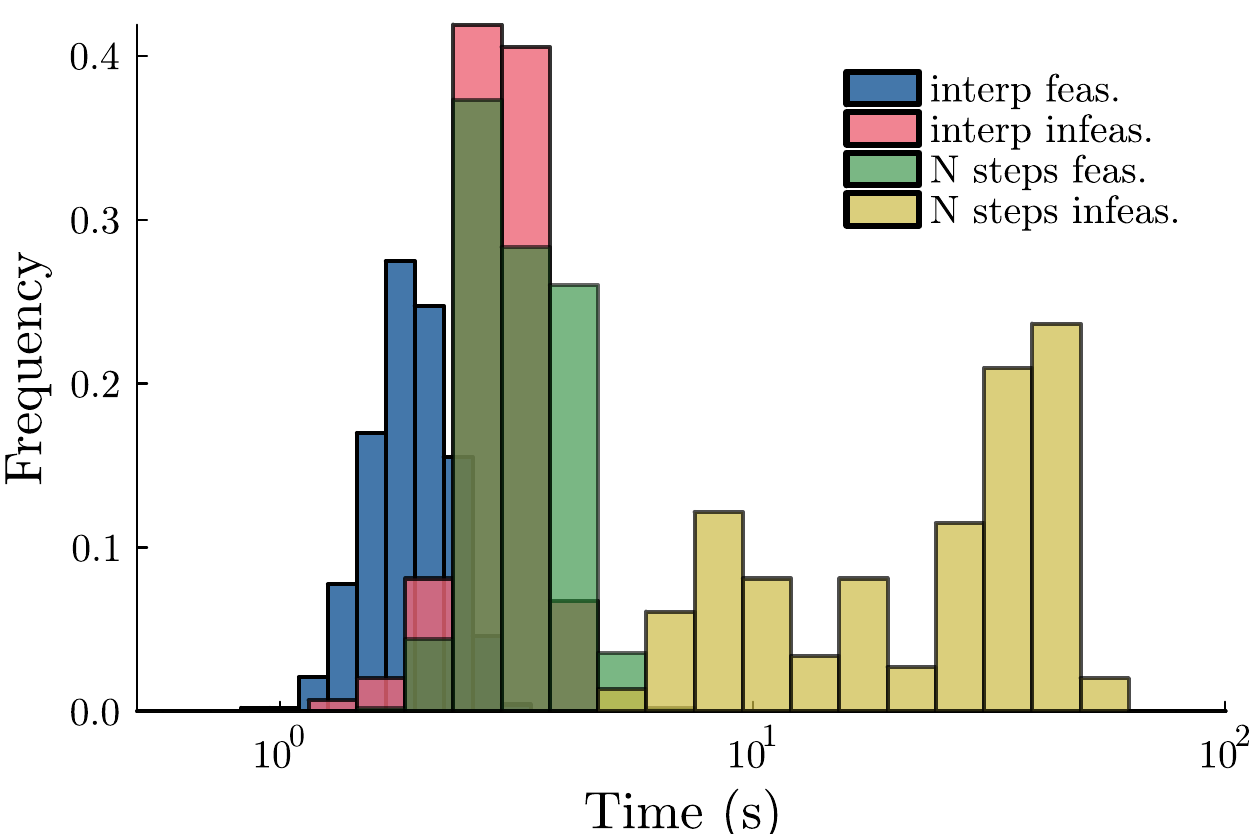}
    \caption{Solve times with the interpolating convex function.}
    \label{cnv-cost-fig: cstr-solve-time-interp}
    \end{subfigure}
    
    \vfill
    \begin{subfigure}{0.95\columnwidth}
    \includegraphics[width=1.\linewidth]{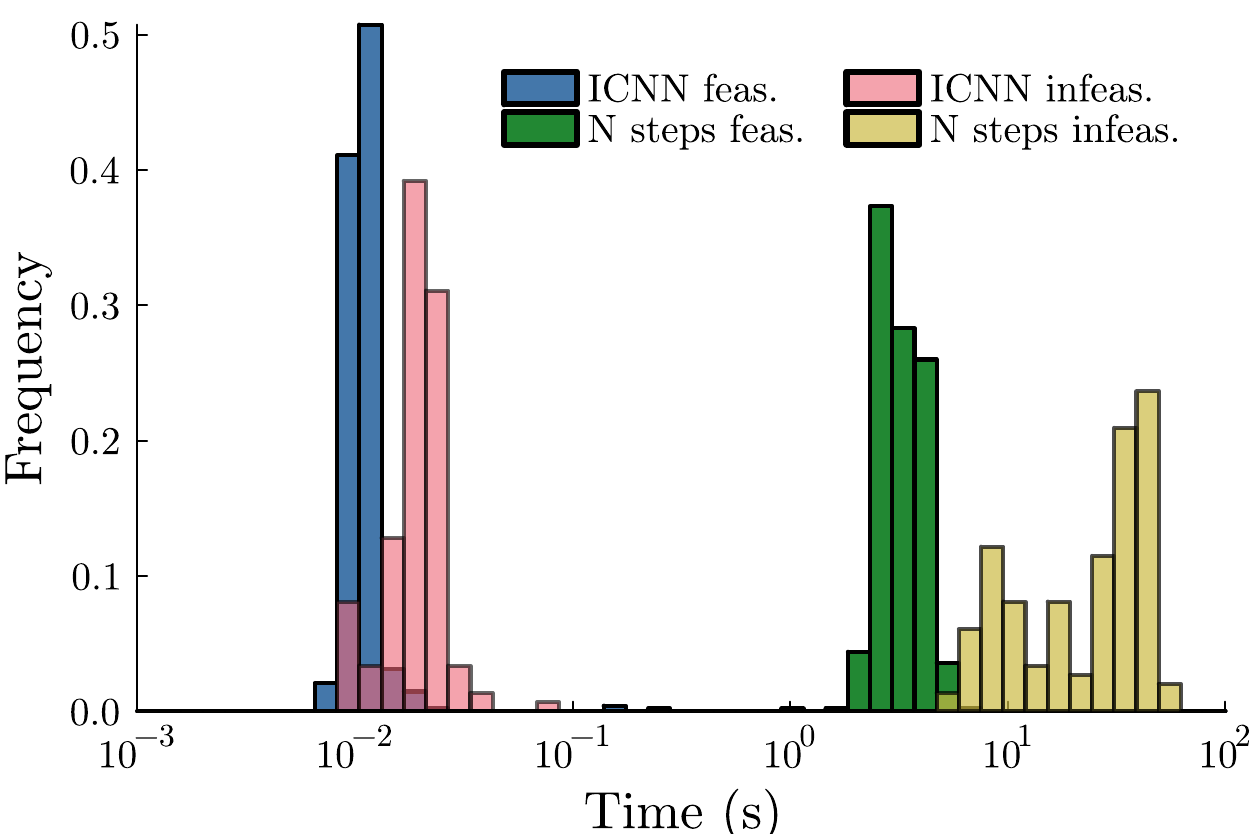}
    \caption{Solve times with the ICNN.}
    \label{cnv-cost-fig: cstr-solve-time-icnn}
    \end{subfigure}
        
    \caption{Solution time using ICNN and interpolating convex functions from feasible and infeasible start points.}
    \label{fig: solve-time-icnn-interp}
\end{figure}

The solution times using these surrogates are shown in Figures \ref{cnv-cost-fig: cstr-solve-time-interp} and \ref{cnv-cost-fig: cstr-solve-time-icnn}. While the ICNN yields a significant speed-up both from feasible and infeasible start points, the interpolating functions only exhibit a minor speed-up from infeasible start points. Thus, the use of ICNNs is preferred over interpolating convex functions based on both computational benefits and better approximation errors. 


\subsubsection{Convexity and value function approximation}

\begin{figure}
    \centering
    \includegraphics[width=.95\linewidth]{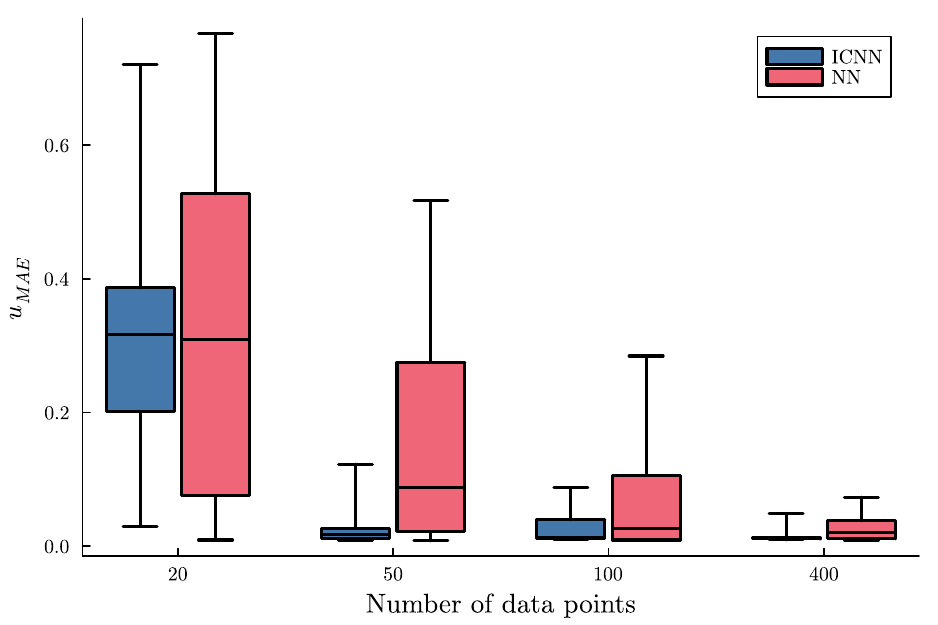}
    \caption{Error in $u$ using ICNN and NN for differing amounts of training data.}
    \label{fig:icnn-vs-nn}
\end{figure}

We compare the difference between using an ICNN and using a standard feed-forward NN based on the data efficiency of the surrogates. To do this, we repeatedly train the networks using differing amounts of training data (randomly selected from all the training data) and evaluate the mean absolute error in the control output when using the trained networks in formulation \eqref{cnv-cost-eq: icnn-mpc-reform}. The results are summarised in Figure \ref{fig:icnn-vs-nn}. Using an ICNN dramatically reduces the variability of the error, with the ICNN showing relatively good performance even with only 50 data points. This is significant for large-scale problems as the generation of training data can be a significant bottleneck due to computational expenses, and sampling densely from a high-dimensional space is not practical (the curse of dimensionality).

\subsubsection{Comparison with policy approximation}

\begin{figure}
    \centering
    \includegraphics[width=.8\linewidth]{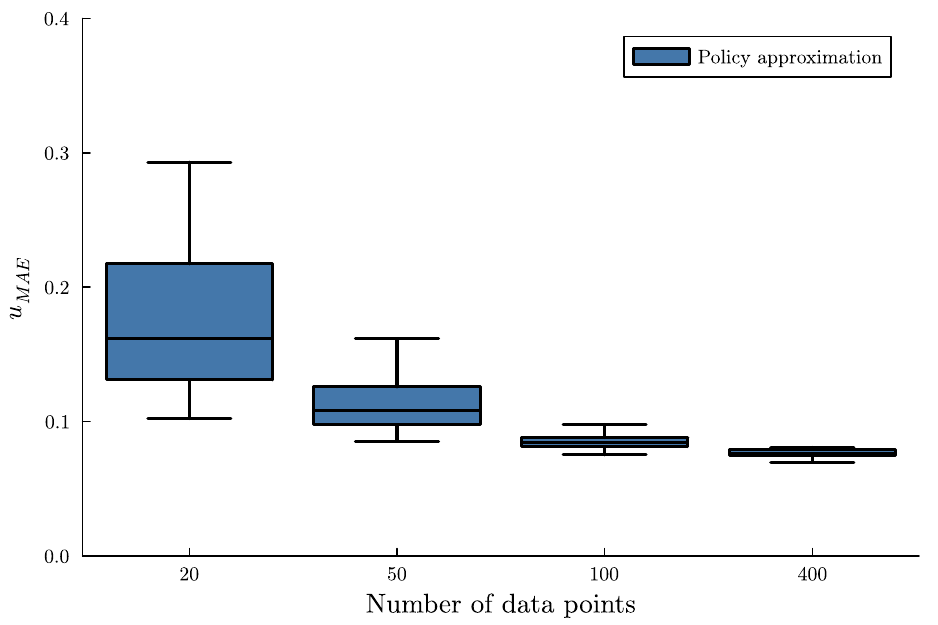}
    \caption{Error in $u$ using an imitation learning formulation with differing amounts of training data.}
    \label{fig:imi-policy-data}
\end{figure}

Lastly, we compare the proposed approach of approximating terms in the objective function with the direct approximation of the control policy defined by the MPC problem. As before, we repeatedly train the policy approximation on different amounts of randomly selected training data and compare its performance using the mean absolute error of the control output. Using the proposed formulation, both a NN and ICNN achieved good performance by 400 data points (Figure \ref{fig:icnn-vs-nn}). However, although the policy approximation initially has better average performance, the approximation still exhibits inferior performance by 400 data points (Figure  \ref{fig:imi-policy-data}). This behaviour is not wholly unexpected. Unlike the proposed approach, where domain knowledge can easily be incorporated into the training and evaluation, it is hard to introduce domain knowledge into the imitation learning problem. This means that the neural network must learn all of the desired behaviour from data, thus the data requirements are expected to be larger. 

\section{Discussion and conclusion}\label{cnv-cost-sec: conc}
This paper introduces a novel approach in which convex objective terms are learned to allow for 1-step ahead MPC problems to be solved when retaining good performance. 
{\color{black}To do so we consider the use of interpolating convex functions and input convex neural networks. We demonstrate the proposed method on two case studies. In both we showed that when using these surrogates the 1-step ahead problem can yeild nearly identical control outputs compared to the original problem. While a significant speed-up is observed in both examples when using an input convex neural network, this is not the case when using the interpolating convex functions. Furthermore, in the first case study we experimentally demonstrated the expected theoretical behaviour of the proposed approach.}  

We also experimentally compared the use of input convex neural networks, with the use of standard feedfoward neural networks to learn objective terms and to directly learn the control policy. The input convex neural networks required significantly less data to achieve good performance. This is important as the computational cost of generating the data to train these surrogates can be prohibitive. We also note that various techniques to improve the data efficiency of training of control policies, e.g. Sobolev training \citep{luken2023sobolev} or data augmentation schemes \citep{krishnamoorthy2021sensitivity}, can be easily applied to the proposed formulation. 


Future work could involve improving the performance of the formulation using the interpolating convex functions. This could be achieved by reducing the number of inequalities, potentially by using a clustering algorithm or other heuristic selection method. Alternatively, other simple convex surrogates could be considered. For example, instead of learning a feasibility surrogate by solving soft constrained problems one could approximate the feasibility surrogate based on the convex hull of the sampled feasible points or can train a convex classifier to learn the boundary of the feasible region, as in \citep{balestriero2022deephull}.

\bibliographystyle{elsarticle-harv} 
\bibliography{bib}





\end{document}